# Study of variations of citation-based parameters for selected Indian science journals

Bidyarthi Dutta

Dept. of Library and Information Science
Vidyasagar University
Rangamati, Midnapore 721 102, West Bengal, India
E-mail: bidyarthi.bhaswati@gmail.com
Mobile: 91-9830884794

# Study of variations of citation-based parameters for selected Indian science journals

**ABSTRACT**

The ratio of the total number of citations to the total number of cited papers was predicted as a constant by Garfield. But, later he observed the changing nature of this constant over time. Scientometricians thus preferred to call it 'Garfield Ratio' rather than 'Garfield Constant'. The 'Garfield Ratio' is a very well-known citation-based parameter, which actually indicates the average citation per cited article. However, Garfield still pointed out that behind this ratio some deeper regularity may be found. In this paper, an analysis of this indicator, the Garfield Ratio $(GR)_Y$ is attempted for twelve distinguished Indian science journals over twelve years' time span ranging from 2009 to 2020. The two other formulae for Garfield Ratio are introduced here, i.e. Modified Garfield Ratio for the year 'Y' $(GRM)_Y$ and Time-Normalised Garfield Ratio for the year 'Y' $(GRN)_Y$. The yearwise analysis of these two Garfield Ratios, i.e. $(GRM)_Y$ and $(GRN)_Y$ for the twelve journals over twelve years are also carried out. The temporal variations of $(GR)_Y$, $(GRM)_Y$ and $(GRN)_Y$ are interpreted by seven statistical parameters, i.e. mean, median, standard deviation, range, coefficient of variation, skewness and kurtosis. The temporal variations of $(GR)_Y$ for twelve journals are viewed by regression analysis to find out the best-fit function.

# INTRODUCTION

*Is the ratio between number of citations and publications cited a true constant?* - this question was raised by Garfield (Garfield, 1977) in 1976. During the compilation of citation indexes of science literature around 1964, Garfield was surprised by the near-constancy in the ratio of 1.7 between references processed each year and the number of different items cited by those references (Garfield, 1977). The name given then to this constant was *Citation Constant*. In Garfield's own words (Garfield, 1977), "If you examine any annual SCI Guide, this 'constant' is readily apparent in the chronological statistical analysis provided. As the number and type of journals covered by SCI has grown, the ratio has changed slightly. Perhaps my own mathematical and statistical naiveté has made it possible for me to suffer in silence so many years while I wondered about the probability of a true constant." Also in 1998 Garfield wrote, ''Another modest contribution I made to the microtheory of citation is Garfield's constant. Actually, we know that this 'constant' is really a ratio. That ratio is remarkably 'stable' considering how much the literature has grown. Due to continuous growth of source journal coverage and increasing references cited per paper, the ratio of citations to published papers increased about 75% from 1945 to 1995—from 1.33 to 2.25 over the past 50 years. It is the inflation of the literature which increases the ratio each year'' (Garfield, 1998). According to Schubert and Schubert (2018), "Although Garfield's Constant has been mentioned a few times in the literature, it was done mainly as a historical curiosity (Bensman, 2007) or just another example of the various laws of bibliometrics (Holden, Rosenberg & Barker, 2005). In the present article, Garfield's idea for twelve selected Indian science journals has been applied.

# LITERATURE REVIEW

The name given to the ratio of the total number of citations to the total number of cited papers was called "Garfield's Constant" in some of the earlier works of Eugene Garfield. As Schubert (Schubert and Schubert, 2018), pointed out later, he himself realized that the ratio is changing over time, but still was confident that behind this ratio some deeper regularity may be found. The ratio of the number of citations per cited item was used as indicators in some other subject areas (Schubert and Schubert, 2018) like psychology (White, 1979 and White & White, 1977), Nordic cancer research (Luukonen-

Gronow and Suutarinen, 1988), Croatian journals (Andreis & Jokic, 2008) and Polish history (Kolasa, 2012). Vinkler (Vinkler, 2007 & 2009) introduced a new indicator, i.e. *Current Contribution Index* (CCI), defined as the ratio of number of citations received by a journal in a given year to the total number of citations obtained by all journals of the corresponding field in that year. He studied the correlation between Garfield Factor and CCI for different journals, but found no significant correlation between them. Umstätter and Nourmohammadi (Umstätter & Hamzeh, 2006) observed the non-constancy of Garfield Constant in *Science Citation Index* over the years. They found that the factors like impact factor, most cited articles and uncitedness played an important role in this context. The Garfield Constant is also known as Garfield Ratio and Garfield Factor. Vinkler indicated Journal Impact Factor of ISI as Garfield (Impact) Factor or Garfield Factor, which is different from this Garfield Ratio or Garfield Constsnt (Vinkler, 2002 & 2004).

## RESEARCH GAP & PURPOSE OF RESEARCH

The ratio between number of citations and publications cited of any subject domain may or may not be a constant, but it reveals numbers of important features on distribution of citation over cited publications. The citation accumulation patterns of the cited items may be foreseen by the analysis of Garfield Constant. No study has ever been found that calculated Garfield Ratio of Indian journals in any subject that certainly creates a research gap in the area of citation analysis of Indian sources. This study intends to bridge the said gap by carrying out the study of Garfield Ratio for twelve distinguished Indian science journals. The original formula for Garfield Ratio is considered for this study. Besides, the original formula is modified and temporally normalised to introduce two new formula for Garfield Ratio, i.e. Modified Garfield Ratio and Normalised Garfield Ratio. The citation behaviours of the twelve stipulated journals are interpreted by analysing all these three Garfield Ratios. The temporal variations of Garfield Ratios for twelve journals are studied by regression analysis to find out the best-fit function.

## SCOPE & METHODOLOGY

The list of the twelve Indian science journals selected for this study is listed below. The abbreviations used for each journal are shown in the adjacent parenthesis:

1) *Defence Science Journal (DSJ)*
2) *Indian Journal of Biochemistry and Biophysics (IJBB)*
3) *Indian Journal of Engineering and Materials Sciences (IJEMS)*
4) *Indian Journal of Physics (IJP)*
5) *Indian Journal of Pure & Applied Physics (IJPAP)*
6) *Journal of Astrophysics and Astronomy (JAA)*
7) *Journal of Earth System Science (JESS)*
8) *Journal of Medical Physics (JMP)*
9) *Journal of Scientific and Industrial Research (JSIR)*
10) *Pramana - Journal of Physics (PJP)*
11) *Proceedings of the Indian National Science Academy (PINSA)*
12) *Proceedings of the National Academy of Sciences India Section A - Physical Sciences (PNASI)*

The primary data have been collected from *Scopus* database. The search strategy followed in *Scopus* under 'Advanced Search' was, "SUBJAREA (PHYS) AND AFFIL COUNTRY (INDIA) AND (EXACTSRCTITLE (DEFENCE SCIENCE JOURNAL))". The period was set from 2009 to 2020. The same strategy was repeated for the other eleven journals as listed above and the total number of papers, number of cited papers [$(CP)_Y$] and total number of citations [$(TC)_Y$] for each of the journals from 2009 to 2020 as retrieved from Scopus are collected. The ages (A) of the papers corresponding to each year are obtained by subtracting the said year from the current year, i.e. 2022. The Garfield Ratio for a particular year (Y) or Yearly Garfield Ratio, denoted by $(GR)_Y$ is defined as the ratio of Total Citation $(TC)_Y$ to the number of Cited Papers $(CP)_Y$ for that particular year (Y). The numerical values of $(GR)_Y$ for twelve said journals since 2009 to 2020 are presented in Table 2, where, $(GR)_Y = (TC)_Y/(CP)_Y$. Two quantities are introduced here, i.e. Fractional Cited Paper $(FC)_Y$ and Fractional Total Citation $(FTC)_Y$ for the year 'Y', which are defined as,

$(FC)_Y = (CP)_Y / \sum (CP)_Y$ and

$(FTC)_Y = (TC)_Y / \sum (TC)_Y$, where

$\sum (CP)_Y = (CP)_{2009} + (CP)_{2010} + (CP)_{2011} + \ldots + (CP)_{2020}$

$\sum (TC)_Y = (TC)_{2009} + (TC)_{2010} + (TC)_{2011} + \ldots + (TC)_{2020}$

The quantities $\sum (CP)_Y$ and $\sum (TC)_Y$ represent the summation of $(CP)_Y$ and $(TC)_Y$ over the entire twelve year time span from 2009 to 2020, for each journal. The name given to the ratio of $\sum (TC)_Y$ to $\sum (CP)_Y$ is Consolidated Garfield Ratio (CGR). The values of CGR for twelve journals are presented in Table 1.

Now, the Modified Garfield Constant for the year 'Y', denoted by $(GRM)_Y$ is defined as,

$(GRM)_Y = (FTC)_Y / (FC)_Y$; The values of $(GRM)_Y$ for twelve journals over twelve years is presented in Table 3.

Also, the Time-Normalised Garfield Constant for the year 'Y', denoted by $(GRN)_Y$ is defined as,

$(GRN)_Y = (GR)_Y / A$, where 'A' indicates age of the papers. The values of $(GRN)_Y$ for twelve journals over twelve years (2009-2020) is presented in Table 4. The Extremal (highest and lowest) values of $(GR)_Y$, $(GRM)_Y$ and $(GRN)_Y$ over twelve years for twelve journals are presented in Table 5. Also, the Extremal values of seven statistical parameters (Mean, Median, Standard Deviation, Range, Coefficient of Variation, Skewness and Kurtosis) over twelve years for twelve journals are presented in Table 6.

### RESULTS & ANALYSIS

As evident from Table 1, the highest CGR (9.01) is observed for *Indian Journal of Biochemistry & Biophysics*, followed by *Journal of Scientific & Industrial Research* (8.57) and *Journal of Earth System Science* (8.54). In terms of total number of citations over twelve years, *Indian Journal of Physics* ranked top (10,714) followed by *Journal of Earth System Science* (10,702) and *Pramana- Journal of Physics* (10,600). Also, in terms of total number of cited papers, *Indian Journal of Physics* ranked top (1771) followed by *Pramana- Journal of Physics* (1647) and *Journal of Earth System Science* (1253). Hence, the articles published in IJBB received maximum citations on average, which indicates its highest citation attraction capacity compared to others. Also, the minimum number of citations is received by the articles published in JAA, PNASI and PINSA, which indicates their least citation attraction capacity. It is found from the analysis of the statistical parameters that they are dependent on the Consolidated Garfield Ratio or CGR, which is a ratio between the sum total of yearly citations to the sum total of yearly number of papers. The CGR thus represents the number of citations received by an individual paper over the entire time span, i.e. twelve years.

The highest and lowest values of the Yearly Garfield Ratio [$(GR)_Y$] are found as 19.27 (for the journal IJBB in 2009) and 1.17 (for the journal JMP in 2020) respectively, which reveals the range of $(GR)_Y$ as 18.1 over twelve years for the entire sample. This fairly large range affirms non-constancy of Garfield Ratio [$(GR)_Y$]. The variational patterns of $(GR)_Y$ for twelve journals over twelve years' time span each are analysed and presented in Table 2A. It is observed that for eight journals the variational patterns followed logarithmic (DSJ, JSIR), polynomial (cubic (IJBB, IJP) and $4^{th}$ degree (IJEMS)), linear (IJPAP, JESS) and exponential (JMP) functions as validated by the corresponding best-fit equations. Also, no exact variational function was observed for four journals (JAA, PJP, PINSA and PNASI) as no befitting equation was found for those and the variation was irregular or haphazard.

The highest and lowest values of $(GRM)_Y$ and $(GRN)_Y$ are found as (2.33, 4.82) and (0.14, 0.22) respectively (Table 5). The ranges of $(GRM)_Y$ and $(GRN)_Y$ are thus 2.19 and 4.6 respectively. Hence, the Modified Garfield Ratio or $(GRM)_Y$ shows least variation compared to $(GRN)_Y$ and $(GR)_Y$. No Garfield Ratio is found here constant, but a variable, though Modified Garfield Ratio $(GRM)_Y$ shows a near-constancy pattern. The $(GR)_Y$ varies for all journals over twelve years (Table 2) with maximum range of variation found in JMP & JSIR (17.74) followed by IJBB (17.1) and JESS (16.14). As the least variation is observed in PNASI (4.58), $(GR)_Y$ is nearly constant in this journal only.

The analysis of statistical parameters (Table 6) shows that the journal JESS holds maximum mean & median values for $(GR)_Y$, $(GRM)_Y$ and $(GRN)_Y$, i.e. 9.4, 1.15 and 1.43 (mean) & 8.69, 1.07 and 1.42

(median) respectively, while the journals PINSA, JSIR & IJBB possess minimum mean and median values. The Lowest Values (LV) of $(GR)_Y$-mean is observed in PINSA (3.52), $(GCM)_Y$-mean is observed in JSIR (0.82) and $(GRN)_Y$-mean is observed in PINSA (0.69). Also, the Lowest Values (LV) of $(GR)_Y$-median is observed in PINSA (3.09), $(GRM)_Y$-median is observed in IJBB (0.61) and $(GRN)_Y$-median is observed in PINSA (0.69). The highest standard deviation values possessed by $(GR)_Y$, $(GRM)_Y$ and $(GRN)_Y$ are 5.91 (IJBB), 0.7 (JESS) & 1.17 (PJP) respectively indicating highest variations in concerned GR values in the corresponding journals shown in parenthesis. Also, the lowest standard deviation values possessed by $(GR)_Y$, $(GRM)_Y$ and $(GRN)_Y$ are 1.47 (PNASI), 0.27 (PJP) & 0.21 (IJPAP) respectively indicating lowest variations or nearly constant pattern in concerned GR values in the corresponding journals shown in parenthesis. However, it is clear from Table 2, Table 3 and Table 4 that the $(GR)_Y$ or Yearly Garfield Ratios are highly varying, though $(GRM)_Y$ or Modified Garfield Constants are moderately varying and $(GRN)_Y$ or Normalised Garfield Constants are sparsely varying. From the analysis of the values of the ranges, it is clear that $(GR)_Y$ varies highest in JMP & JSIR and lowest in PNASI. Similarly, the highest & lowest variations of $(GRM)_Y$ & $(GRN)_Y$ are observed in (JMP, PJP) and in (PJP, IJPAP) respectively. It is interesting to note that, the journal PJP shows lowest variation or almost constancy pattern in $(GRM)_Y$, whereas highest variation in $(GRN)_Y$ at the same time.

The analysis of the Coefficients of Variation shows that the highest values for $(GR)_Y$, $(GRM)_Y$ and $(GRN)_Y$ are found in JSIR (0.78) and PNASI (0.86) respectively. The lowest Coefficients of Variations for $(GR)_Y$, $(GRM)_Y$ and $(GRN)_Y$ are found in PJP (0.27) and IJPAP (0.2) respectively. The temporal variations of $(GR)_Y$ for all journals are positively skewed except *Indian Journal of Physics*, which is negatively skewed (Table 2). Also, the temporal variations of $(GRM)_Y$ for all journals are positively skewed (except *Indian Journal of Physics*) and the same for $(GRN)_Y$ are positively skewed for all journals. The analysis of kurtosis for $(GR)_Y$ and $(GRM)_Y$ shows (Table 2 & Table 3), nine journals have negative kurtosis values except JMP, JSIR & PINSA, which have positive values. But, the kurtosis analysis for $(GRN)_Y$ (Table 4) shows eleven journals having positive kurtosis values except *Journal of Astrophysics and Astronomy*, which has negative value close to zero. The temporal distribution of (GRN)$_Y$ for *Journal of Astrophysics and Astronomy* thus shows almost symmetrical or normal distribution pattern. All kurtosis values for (GR)$_Y$ and (GRM)$_Y$ are found less than 3, indicating platykurtic distribution having thin tails than normal distribution with fewer extreme positive or negative values. But for (GRN)$_Y$, the kurtosis values for only 6 journals have greater than 3 (leptokurtic distribution) and for 6 journals have less than 3 (platykurtic distribution). The journals IJBB, IJEMS, IJP, IJPAP, PJP and PINSA show leptokurtic distribution over entire time span, which have wide or thick tails having larger number of extreme positive or negative values. Actually, the values of (GRN)$_Y$ for these journals in 2020 are far greater than other values, which resulted high positive kurtosis values. Hence, the distributions of (GR)$_Y$ and (GRM)$_Y$ are more or less alike, while the distribution of (GRN)$_Y$ is fairly different.

Table 1: The Consolidated Garfield Ratios (CGR) for twelve journals

| The Journals (Indian physics and astronomy journals) | Total number of cited papers over 12 years (2009-2020) [∑(CP)$_Y$] | Total Citation since 2009 to 2020 [∑(TC)$_Y$] | CGR = [∑(TC)$_Y$/ ∑(CP)$_Y$] |
|---|---|---|---|
| Indian Journal of Biochemistry and Biophysics | 580 | 5227 | 9.01 |
| Journal of Scientific and Industrial Research | 789 | 6761 | 8.57 |
| Journal of Earth System Science | 1253 | 10702 | 8.54 |
| Journal of Medical Physics | 340 | 2763 | 8.13 |
| Indian Journal of Pure and Applied Physics | 977 | 7296 | 7.47 |
| Defence Science Journal | 574 | 4046 | 7.05 |
| Indian Journal of Engineering and Materials Sciences | 542 | 3814 | 7.04 |
| Pramana- Journal of Physics | 1647 | 10600 | 6.44 |
| Indian Journal of Physics | 1771 | 10714 | 6.05 |
| Journal of Astrophysics and Astronomy | 363 | 1763 | 4.86 |
| Proceedings of the National Academy of Sciences India Section A - Physical Sciences | 458 | 2128 | 4.65 |
| Proceedings of the Indian National Science Academy | 383 | 1551 | 4.05 |

Table 2: Yearwise variation of Garfield Ratio (GR)_Y for twelve journals

| Journals | 2009 | 2010 | 2011 | 2012 | 2013 | 2014 | 2015 | 2016 | 2017 | 2018 | 2019 | 2020 | Mean | Median | Standard Deviation | Range | Coefficient of Variation | Skewness | Kurtosis |
|---|---|---|---|---|---|---|---|---|---|---|---|---|---|---|---|---|---|---|---|
| DSJ | 14.35 | 13.48 | 9.70 | 8.61 | 6.57 | 6.64 | 4.54 | 6.03 | 3.83 | 3.71 | 2.03 | 1.56 | 6.75 | 6.30 | 4.13 | 12.79 | 0.61 | 0.72 | -0.31 |
| IJBB | 19.27 | 15.66 | 13.07 | 10.45 | 9.82 | 6.51 | 4.48 | 2.17 | 2.64 | 2.92 | 2.25 | 2.65 | 7.66 | 5.50 | 5.91 | 17.1 | 0.77 | 0.82 | -0.56 |
| IJEMS | 7.57 | 11.66 | 11.43 | 8.36 | 8.92 | 5.63 | 4.96 | 5.57 | 3.10 | 2.90 | 1.45 | 1.73 | 6.11 | 5.60 | 3.52 | 10.21 | 0.58 | 0.28 | -1.09 |
| IJP | 6.76 | 7.46 | 7.69 | 7.25 | 8.58 | 7.26 | 5.36 | 5.17 | 4.84 | 4.75 | 2.80 | 3.35 | 5.94 | 6.06 | 1.82 | 5.78 | 0.31 | -0.35 | -0.96 |
| IJPAP | 9.58 | 12.25 | 9.16 | 7.91 | 6.86 | 6.83 | 5.55 | 5.11 | 3.86 | 3.29 | 2.43 | 1.58 | 6.20 | 6.19 | 3.19 | 10.67 | 0.51 | 0.34 | -0.49 |
| JAA | 6.08 | 3.91 | 6.43 | 5.58 | 8.75 | 3.31 | 4.64 | 3.29 | 6.49 | 2.87 | 2.55 | 1.86 | 4.65 | 4.28 | 2.04 | 6.89 | 0.44 | 0.55 | -0.35 |
| JESS | 18.11 | 17.31 | 14.00 | 14.16 | 12.96 | 8.68 | 8.69 | 5.96 | 5.55 | 3.13 | 1.97 | 2.22 | 9.40 | 8.69 | 5.79 | 16.14 | 0.62 | 0.16 | -1.45 |
| JMP | 11.61 | 18.91 | 9.13 | 9.41 | 9.00 | 7.50 | 4.97 | 5.55 | 4.47 | 2.00 | 2.80 | 1.17 | 7.21 | 6.53 | 4.92 | 17.74 | 0.68 | 1.12 | 1.80 |
| JSIR | 19.82 | 15.23 | 9.02 | 7.34 | 7.29 | 6.82 | 4.10 | 3.39 | 4.00 | 3.07 | 2.08 | 2.19 | 7.03 | 5.46 | 5.47 | 17.74 | 0.78 | 1.50 | 1.76 |
| PJP | 8.94 | 7.12 | 7.54 | 8.29 | 8.93 | 4.34 | 5.71 | 5.45 | 5.93 | 4.65 | 4.44 | 4.82 | 6.35 | 5.82 | 1.74 | 4.6 | 0.27 | 0.40 | -1.47 |
| PINSA | 3.40 | 5.50 | 2.17 | 5.90 | 2.00 | 7.92 | 3.67 | 3.56 | 2.78 | 1.96 | 1.53 | 1.81 | 3.52 | 3.09 | 1.98 | 6.39 | 0.56 | 1.18 | 0.73 |
| PNASI | 2.78 | 3.54 | 2.60 | 5.61 | 2.60 | 7.18 | 5.04 | 5.59 | 4.35 | 4.10 | 5.50 | 3.07 | 4.33 | 4.23 | 1.47 | 4.58 | 0.34 | 0.43 | -0.69 |

(The abbreviations of the journal names are used here and in subsequent Tables)

Table 2A: Best-fit functions representing the yearwise variations of Garfield Ratio (GR)_Y

| Journals | Best-Fit Function | Best-Fit Equation | Values of the Constants & Coeff. of Determination ($R^2$) | | | | | |
|---|---|---|---|---|---|---|---|---|
| | | | a | b | c | d | e | $R^2$ |
| DSJ | Logarithmic | $Y = a*\ln(X) + b$ | -5.332 | 15.635 | --- | --- | --- | 0.953 |
| IJBB | Non-Linear (Cubic Polynomial) | $Y = a(X)^3 + b(X)^2 + c(X) + d$ | 0.008 | 0.027 | 3.034 | 21.976 | --- | 0.988 |
| IJEMS | Non-Linear (4th Deg. Polynomial) | $Y = a(X)^4 + b(X)^3 + c(X)^2 + d(X) + e$ | -0.007 | 0.216 | -2.198 | 7.642 | 2.500 | 0.934 |
| IJP | Non-Linear (Cubic Polynomial) | $Y = a(X)^3 + b(X)^2 + c(X) + d$ | 0.013 | -0.311 | 1.667 | 5.300 | --- | 0.903 |
| IJPAP | Linear | $Y = a + b(X)$ | -0.854 | 11.752 | --- | --- | --- | 0.934 |
| JAA | Irregular & no befitting equation | --- | --- | --- | --- | --- | --- | --- |
| JESS | Linear | $Y = a + b(X)$ | -1.584 | 19.691 | --- | --- | --- | 0.973 |
| JMP | Exponential | $Y = a*e^{bX}$ or $Y = a*\exp(bX)$ | 21.380 | -0.210 | --- | --- | --- | 0.873 |
| JSIR | Logarithmic | $Y = a*\ln(X) + b$ | -7.096 | 18.849 | --- | --- | --- | --- |
| PJP | Irregular & no befitting equation | --- | --- | --- | --- | --- | --- | --- |
| PINSA | Irregular & no befitting equation | --- | --- | --- | --- | --- | --- | --- |
| PNASI | Irregular & no befitting equation | --- | --- | --- | --- | --- | --- | --- |

Table 3: Yearwise variation of Modified Garfield Ratio (GRM)_Y for twelve journals

| Journals | 2009 | 2010 | 2011 | 2012 | 2013 | 2014 | 2015 | 2016 | 2017 | 2018 | 2019 | 2020 | Mean | Median | Standard Deviation | Range | Coefficient of Variation | Skewness | Kurtosis |
|---|---|---|---|---|---|---|---|---|---|---|---|---|---|---|---|---|---|---|---|
| DSJ | 2.04 | 1.91 | 1.38 | 1.22 | 0.93 | 0.94 | 0.64 | 0.86 | 0.54 | 0.53 | 0.29 | 0.22 | 0.96 | 0.89 | 0.59 | 1.81 | 0.61 | 0.72 | -0.31 |
| IJBB | 2.14 | 1.74 | 1.45 | 1.16 | 1.09 | 0.72 | 0.50 | 0.24 | 0.29 | 0.32 | 0.25 | 0.29 | 0.85 | 0.61 | 0.66 | 1.90 | 0.77 | 0.82 | -0.56 |
| IJEMS | 1.08 | 1.66 | 1.62 | 1.19 | 1.27 | 0.80 | 0.70 | 0.79 | 0.44 | 0.41 | 0.21 | 0.25 | 0.87 | 0.80 | 0.50 | 1.45 | 0.58 | 0.28 | -1.09 |
| IJP | 1.12 | 1.23 | 1.27 | 1.20 | 1.42 | 1.20 | 0.89 | 0.85 | 0.80 | 0.78 | 0.46 | 0.55 | 0.98 | 1.00 | 0.30 | 0.96 | 0.31 | -0.35 | -0.96 |
| IJPAP | 1.28 | 1.64 | 1.23 | 1.06 | 0.92 | 0.91 | 0.74 | 0.68 | 0.52 | 0.44 | 0.33 | 0.21 | 0.83 | 0.83 | 0.43 | 1.43 | 0.51 | 0.34 | -0.49 |
| JAA | 1.25 | 0.80 | 1.32 | 1.15 | 1.80 | 0.68 | 0.96 | 0.68 | 1.34 | 0.59 | 0.52 | 0.38 | 0.96 | 0.88 | 0.42 | 1.42 | 0.44 | 0.55 | -0.35 |
| JESS | 2.12 | 2.13 | 1.72 | 1.74 | 1.59 | 1.07 | 1.07 | 0.73 | 0.68 | 0.38 | 0.24 | 0.27 | 1.15 | 1.07 | 0.70 | 1.89 | 0.61 | 0.11 | -1.53 |
| JMP | 1.43 | 2.33 | 1.12 | 1.16 | 1.11 | 0.92 | 0.61 | 0.68 | 0.55 | 0.25 | 0.34 | 0.14 | 0.89 | 0.80 | 0.61 | 2.18 | 0.68 | 1.12 | 1.80 |
| JSIR | 2.31 | 1.78 | 1.05 | 0.86 | 0.85 | 0.80 | 0.48 | 0.40 | 0.47 | 0.36 | 0.24 | 0.26 | 0.82 | 0.64 | 0.64 | 2.07 | 0.78 | 1.50 | 1.76 |
| PJP | 1.39 | 1.11 | 1.17 | 1.29 | 1.39 | 0.67 | 0.89 | 0.85 | 0.92 | 0.72 | 0.69 | 0.75 | 0.99 | 0.90 | 0.27 | 0.71 | 0.27 | 0.40 | -1.47 |
| PINSA | 0.84 | 1.36 | 0.54 | 1.46 | 0.49 | 1.95 | 0.91 | 0.88 | 0.69 | 0.48 | 0.38 | 0.45 | 0.87 | 0.76 | 0.49 | 1.58 | 0.56 | 1.18 | 0.73 |
| PNASI | 0.60 | 0.76 | 0.56 | 1.21 | 0.56 | 1.55 | 1.09 | 1.20 | 0.94 | 0.88 | 1.18 | 0.66 | 0.93 | 0.91 | 0.32 | 0.99 | 0.34 | 0.43 | -0.69 |

Table 4: Yearwise variation of Time-Normalised Garfield Ratio $(GRN)_Y$ for twelve journals

| Journals | Years Ranging from 2009 to 2020 | | | | | | | | | | | | Statistical Parameters | | | | | | |
|---|---|---|---|---|---|---|---|---|---|---|---|---|---|---|---|---|---|---|---|
| | 2009 | 2010 | 2011 | 2012 | 2013 | 2014 | 2015 | 2016 | 2017 | 2018 | 2019 | 2020 | Mean | Median | Standard Deviation | Range | Coefficient of Variation | Skewness | Kurtosis |
| DSJ | 1.20 | 1.23 | 0.97 | 0.96 | 0.82 | 0.95 | 0.76 | 1.21 | 0.96 | 1.24 | 1.01 | 1.56 | 1.07 | 0.99 | 0.22 | 0.80 | 0.21 | 0.78 | 0.78 |
| IJBB | 1.61 | 1.42 | 1.31 | 1.16 | 1.23 | 0.93 | 0.75 | 0.43 | 0.66 | 0.97 | 1.13 | 2.65 | 1.19 | 1.14 | 0.57 | 2.22 | 0.48 | 1.52 | 3.67 |
| IJEMS | 0.63 | 1.06 | 1.14 | 0.93 | 1.12 | 0.80 | 0.83 | 1.11 | 0.78 | 0.97 | 0.73 | 1.73 | 0.99 | 0.95 | 0.29 | 1.10 | 0.29 | 1.55 | 3.56 |
| IJP | 0.56 | 0.68 | 0.77 | 0.81 | 1.07 | 1.04 | 0.89 | 1.03 | 1.21 | 1.58 | 1.40 | 3.35 | 1.20 | 1.04 | 0.74 | 2.79 | 0.62 | 2.56 | 7.50 |
| IJPAP | 0.80 | 1.11 | 0.92 | 0.88 | 0.86 | 0.98 | 0.93 | 1.02 | 0.97 | 1.10 | 1.21 | 1.58 | 1.03 | 0.97 | 0.21 | 0.78 | 0.20 | 1.80 | 3.97 |
| JAA | 0.51 | 0.36 | 0.64 | 0.62 | 1.09 | 0.47 | 0.77 | 0.66 | 1.62 | 0.96 | 1.27 | 1.86 | 0.90 | 0.72 | 0.47 | 1.50 | 0.53 | 0.96 | -0.03 |
| JESS | 1.51 | 1.57 | 1.40 | 1.57 | 1.62 | 1.24 | 1.45 | 1.19 | 1.39 | 1.04 | 0.99 | 2.22 | 1.43 | 1.42 | 0.32 | 1.23 | 0.23 | 1.06 | 2.49 |
| JMP | 0.97 | 1.72 | 0.91 | 1.05 | 1.13 | 1.07 | 0.83 | 1.11 | 1.12 | 0.67 | 1.40 | 1.17 | 1.09 | 1.09 | 0.27 | 1.05 | 0.25 | 0.94 | 2.01 |
| JSIR | 1.65 | 1.38 | 0.90 | 0.82 | 0.91 | 0.97 | 0.68 | 0.68 | 1.00 | 1.02 | 1.04 | 2.19 | 1.10 | 0.99 | 0.44 | 1.51 | 1.63 | 2.58 | 0.40 |
| PJP | 0.74 | 0.65 | 0.75 | 0.92 | 1.12 | 0.62 | 0.95 | 1.09 | 1.48 | 1.55 | 2.22 | 4.82 | 1.41 | 1.02 | 1.17 | 4.20 | 0.83 | 2.62 | 7.52 |
| PINSA | 0.28 | 0.50 | 0.22 | 0.66 | 0.25 | 1.13 | 0.61 | 0.71 | 0.69 | 0.65 | 0.77 | 1.81 | 0.69 | 0.65 | 0.44 | 1.59 | 0.63 | 1.62 | 3.55 |
| PNASI | 0.23 | 0.32 | 0.26 | 0.62 | 0.33 | 1.03 | 0.84 | 1.12 | 1.09 | 1.37 | 2.75 | 3.07 | 1.08 | 0.93 | 0.94 | 2.84 | 0.86 | 1.37 | 1.08 |

Table 5: Extremal values and Ranges of different Garfield Ratios

| | Highest Values | | | Lowest Values | | | |
|---|---|---|---|---|---|---|---|
| | Journal | Year | Value | Journal | Year | Value | Range |
| Garfield Ratio $(GR)_Y$ | IJBB | 2009 | 19.27 | JMP | 2020 | 1.17 | 18.1 |
| Modified Garfield Ratio $(GRM)_Y$ | JMP | 2010 | 2.33 | JMP | 2020 | 0.14 | 2.19 |
| Time-Normalised Garfield Ratio $(GRN)_Y$ | PJP | 2020 | 4.82 | PINSA | 2011 | 0.22 | 4.6 |

Table 6: Extremal values of the statistical parameters of different Garfield Ratios

| | Mean | | Median | | Standard Deviation | | Range | | Coeff. of Variation | | Skewness | | Kurtosis | |
|---|---|---|---|---|---|---|---|---|---|---|---|---|---|---|
| | HV | LV | HV | LV | HV | LV | HV | LV | HV | LV | HV | LV | HV | LV |
| $(GC)_Y$ | JESS (9.4) | PINSA (3.52) | JESS (8.69) | PINSA (3.09) | IJBB (5.91) | PNASI (1.47) | JMP & JSIR (17.74) | PNASI (4.58) | JSIR (0.78) | PJP (0.27) | JSIR (1.5) | IJP (-0.35) | JMP (1.8) | PJP (-1.47) |
| $(GCM)_Y$ | JESS (1.15) | JSIR (0.82) | JESS (1.07) | IJBB (0.61) | JESS (0.7) | PJP (0.27) | JMP (2.18) | PJP (0.71) | JSIR (0.78) | PJP (0.27) | JSIR (1.5) | IJP (-0.35) | JMP (1.8) | JESS (-1.53) |
| $(GCN)_Y$ | JESS (1.43) | PINSA (0.69) | JESS (1.42) | PINSA (0.69) | PJP (1.17) | IJPAP (0.21) | PJP (4.2) | IJPAP (0.78) | PNASI (0.86) | IJPAP (0.2) | PJP (2.62) | DSJ (0.78) | PJP (7.52) | JAA (-0.03) |

[HV = Highest Value; LV = Lowest Value]

## CONCLUSION & FUTURE DIRECTION

As Schubert said (Schubert & Schubert, 2018), "The Garfield Constant was changing in time, as Garfield himself admitted, and preferred to call it a ratio". In this paper it is shown that the Garfield Constant or Garfield Ratio, or the ratio of citations per cited papers, is not a constant, but a variable for twelve esteemed Indian physics and astrophysics journals since 2009 to 2020. The two derived formula for Garfield Ratios are also formulated here, i.e. Modified Garfield Ratio and Time-Normalised Garfield Ratio, but they are also not constants due to non-constancy of the original Garfield Ratio. However, Schubert (Schubert & Schubert, 2018) opined to call it as a simple indicator that can be usefully applied in various levels of scientometric analysis. But the fact is that, whether constant or variable, Garfield Ratio may be a very good citation indicator as it averages the number of citations per cited paper. This indicator may be partitioned with respect to number of citations. For instance, for citation ranges 1-5, or 5-10, or 10-50 and so on. Different citation zones may be defined in accordance with different ranges, i.e. low-citation zone, medium-citation zone and high-citation zone as citation-distribution pattern over the publications do not follow a simple average model but highly skewed distribution. The values of Garfield Ratios in different zones for any subject domain or journal may be found out and subsequently compared. It is also important to find out the variational patterns of Garfield Ratio for different cited items over different temporal boundaries to understand how the citation accumulation by different cited items changes over time. The concept of this indicator "Garfield Ratio" should be further studied to probe into the accurate pattern of variation of citations.

## ACKNOWLEDGEMENT

This work is executed under the research project entitled *Design and development of comprehensive database and scientometric study of Indian research output in physics and space science since independence* sponsored by Department of Science and Technology, Govt. of India under NSTMIS scheme, (Vide F. No. DST/NSTMIS/05/252/2017-18 dated 11/01/2018).